\documentclass[twocolumn]{ijns}
\usepackage{graphicx,epsfig,color}
\usepackage[normal]{subfigure}
\usepackage{amsmath,amsthm,amssymb}
\usepackage{longtable}
\usepackage{multirow}
\def\markboth#1#2{\def\leftmark{#1}\def\rightmark{#2}}

\usepackage[normalem]{ulem}
\newcommand\suppress[1]{}

\newlength\wvtextpercent
\author{Biju Issac\\
\normalsize Information and Security Research (iSECURES) Lab, Swinburne University of Technology (Sarawak Campus)\\
\normalsize Kuching, Sarawak, Malaysia (Email: bissac@swinburne.edu.my)\\
{\it\normalsize (Received Apr. 24, 2007; revised and accepted Aug. 31, 2007 \& Jan. 30, 2008)} 
\medskip ~\\
 }
\title {\huge\bf{Secure ARP and Secure DHCP Protocols to Mitigate Security Attacks}}

\date
\newpage
\begin{document}
\maketitle
\thispagestyle{headings}

\begin{abstract}
For network computers to communicate to one another, they need to know one another's IP address and MAC address. Address Resolution Protocol (ARP) is developed to find the Ethernet address that map to a specific IP address. The source computer broadcasts the request for Ethernet address and eventually the target computer replies. The IP to Ethernet address mapping would later be stored in an ARP Cache for some time duration, after which the process is repeated. Since ARP is susceptible to ARP poisoning attacks, we propose to make it unicast, centralized and secure, along with a secure design of DHCP protocol to mitigate MAC spoofing. The secure protocol designs are explained in detail. Lastly we also discuss some performance issues to show how the proposed protocols work.
\end{abstract}

\section{Introduction}

The data link layer hardware does not understand the IP addresses. It only understands the physical address or MAC address. A computer cannot use MAC address alone to communicate to others in a network. Usually the computers are attached to any network using a network interface card that has with a unique physical address called as MAC address (or 48-bit Ethernet address). No two cards would have the same address since such network card manufacturers get the card numbers from a central authority that would assign only unique MAC addresses. This can very well avoid MAC address conflict. These cards know nothing about the IP address of the computer where it is housed [1].  In the following sections we outline two new protocols -- a new centralized protocol called Secure Unicast Address Resolution Protocol (S-UARP) to mitigate ARP poisoning attacks and a new secure DHCP protocol to mitigate MAC spoofing attacks. The organization of further sections are as follows: Section 2 describes on the current ARP and DHCP protocols, section 3 briefly explains on the ARP poisoning and its implementation, section 4 shows related work, section 5 to 7 explains on the S-UARP protocol proposal and related issues, section 8 is on MAC spoofing attack and implementation, section 9 is on secure DHCP protocol, section 10 is performance analysis and section 11 is the conclusion.

\section{The ARP and DHCP Protocols}
\subsection{The Address Resolution Protocol (ARP)}

Address Resolution Protocol (ARP) which is defined in RFC 826 [2] is used to map the IP addresses onto the data link layer MAC address. It is explained as follows. Consider the Figure 1 on interconnected networks. 

\begin{figure}[!h]
\includegraphics[width=\columnwidth]{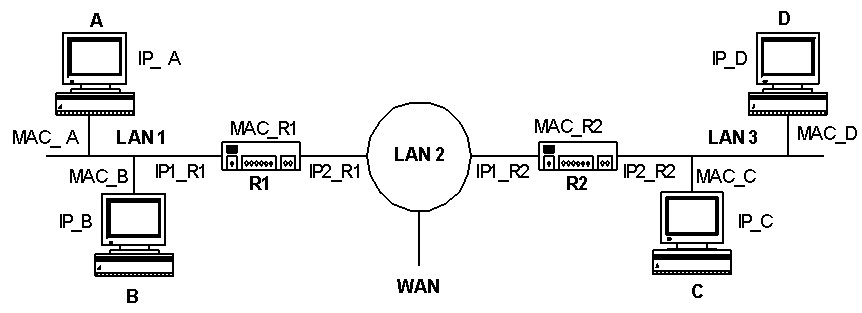}
\caption{Interconnected networks (two bus networks with a ring network in middle).  Each computer on LAN has been shown with an IP address and MAC address.
}\label{fig:01}
\end{figure}

We note that two computers (A and B) on LAN1 have IP address, MAC address pair as [IP\_A, MAC\_A] and [IP\_B, MAC\_B] respectively. Similarly, two computers on LAN3 (C and D) have IP address, MAC address pair as [IP\_C, MAC\_C], [IP\_D, MAC\_D].  Also note that the two routers (R1 and R2) between the networks have two IP address corresponding to the link to bus and ring network. Each router possesses unique MAC address. For a user A on LAN1 to send packets to user on B within LAN1 the following happens:  A query to DNS would return the IP address IP\_B. It then frames a packet with IP\_B in the destination field and passes it to IP layer to transmit. The IP layer sees that the address is on the same network. But it needs to find B's MAC address. To find that it broadcasts a packet asking, ``Who own IP address IP\_B?''. This broadcast would reach on all computers in LAN1. Only computer B would respond with its MAC address MAC\_B.  Thus ARP works by this request and reply approach. The method is quite simple [1].

Some optimizations are possible with ARP. Once computer A gets the ARP reply from B, it stores that IP-to-MAC address mapping of B in a local cache. So if in a short period of time, if A wants to communicate with B, it refers to the local ARP cache, eliminating a second broadcast. Usually, A would include its IP-to-MAC address mapping in the ARP packet, thus informing B of its mapping. In fact all machines on LAN1 can enter this mapping information on A into their ARP cache. Another optimization is to have every computer broadcast its mapping when it boots, in the form of an ARP looking for its won address. To allow for changes in mapping, especially when network card breaks down, and is replaced with a new one, entries in ARP cache should time out after few minutes [3].

\subsection{The Dynamic Host Configuration Protocol (DHCP)}
DHCP stands for `Dynamic Host Configuration Protocol' and is a way by which networked computers get their TCP/IP networking settings from a central server. Dynamic Host Control Protocol (DHCP) is defined in RFC 2131 [4] and 2132 [5]. It is an extension of BOOTP, the previous IP allocation specification. It allows manual and dynamic IP address assignment to computers that requests for that. DHCP server is not reachable by broadcasting from a different network. Hence a \textit{DHCP relay agent} is needed to forward the DHCP DISCOVER broadcast packet from a newly booted machine. It is send as a unicast transmission to the DHCP server (which may be on another network) by the relay agent. The relay agent usually keeps the IP address of the DHCP server. Thus the relay agent is for relaying packets between servers and clients. This makes the DHCP server handle the sub-net that has no server available and thus there is no need to setup a server per sub-net. To keep track of the duration of IP address assignment, a DHCP server uses the concept of \textit{leasing}. As mentioned before, the DHCP server assigns IP addresses automatically from a pool of IP addresses. If a compute leaves the network `abruptly' and does not return the IP address that it was using, that IP address is lost for any further assignment. As a precaution to that, assignment of IP address is only for a fixed duration of time, called leasing. Just before the expiry of the lease, a computer should request the DHCP server for renewal. Otherwise, that IP address cannot be used further [1].

\begin{figure}
\includegraphics[width=\columnwidth]{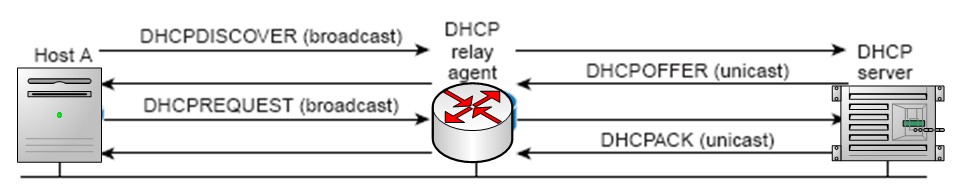}	
\caption{DHCP protocol operation}\label{fig:2}
\end{figure}

A DHCP client may receive offers from multiple DHCP servers and can accept any one of the offers; however, the client usually accepts the first offer it receives. Additionally, the offer from the DHCP server is not a guarantee that the IP address will be allocated to the client; however, the server usually reserves the address until the client has had a chance to formally request the address. The client returns a formal request for the offered IP address to the DHCP server in a DHCPREQUEST broadcast message. The DHCP server confirms that the IP address has been allocated to the client by returning a DHCPACK unicast message to the client as in Figure 2.

The formal request for the offered IP address (the DHCPREQUEST message) that is sent by the client is broadcast so that all other DHCP servers that received the DHCPDISCOVER broadcast message from the client can reclaim the IP addresses that they offered to the client.

If the configuration parameters sent to the client in the DHCPOFFER unicast message by the DHCP server are invalid (a misconfiguration error exists), the client returns a DHCPDECLINE broadcast message to the DHCP server.

The DHCP server will send to the client a DHCPNAK denial broadcast message, which means the offered configuration parameters have not been assigned, if an error has occurred during the negotiation of the parameters or the client has been slow in responding to the DHCPOFFER message (the DHCP server assigned the parameters to another client) of the DHCP server.

\section{ARP Poisoning Security Attack and Implementation}
In ARP Poisoning, forged ARP request and reply packets are used to update the target computer's ARP cache. The target computer is being fooled into believing that the attacker computer (which has a totally different MAC and IP address) as the computer that has the desired IP address with a specific MAC address.  Thus, the attacker can monitor the packet sent by the target computer to the original destination since it is sent to the attacker's computer first before they are sent to the original destination [6]. 

In the ARP poisoning experiment, two desktop computers and one laptop was used as in Figure 3. The two desktop computers (Computer A and Computer B) acted as the victims while the laptop (Computer C) acted as the attacker as in Figure 4. A was the source while B was the destination. C was equipped with the Ethereal packet capturing software [7] and the ARP poisoning software known as Cain \& Abel [8]. Computer A was used to send continuous ICMP packets to B by pinging B. When ARP poisoning was carried out using Cain and Abel software installed on C on Computer A's ARP cache, it was observed in the Ethereal software on C that the ICMP packets were sent only between Computer A and Computer C (attacker), even though A sent it to B. In Cain and Abel software, it was observed that Computer C could monitor the ICMP packets sent between those two computers. It showed that Computer A has been fooled to send ICMP packets to Computer C, which has a different set of MAC and IP address from Computer B. Also, Computer C could then forward these packets to B, after keeping a copy to itself.

\begin{figure}
\centering
	\includegraphics[width=7cm]{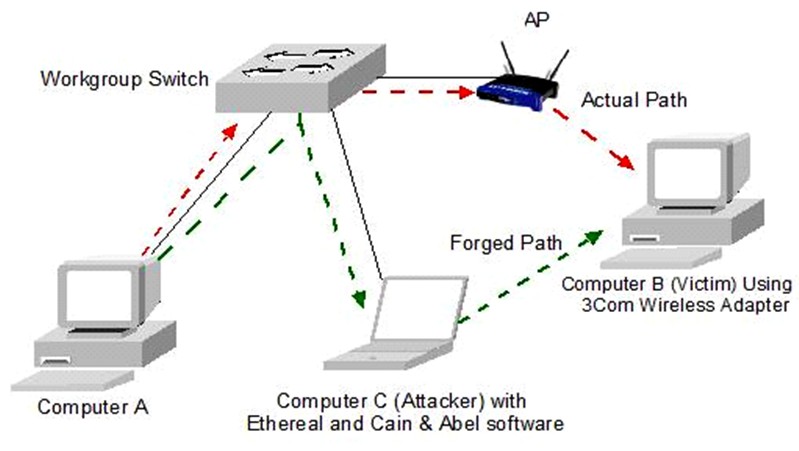}
	\caption{ARP Poisoning Implementation in our lab.}\label{Fig:3}
\end{figure}
Thus ARP poisoning is a method of attacking a network by updating the target computer's ARP cache with forged ARP request and reply packets in an effort to change the Layer 2 Ethernet MAC address to one that the attacker can monitor. The target computer sends frames that were meant for the original destination to the attacker's computer, so that the frames can be read since the ARP replies have been forged. A successful ARP attempt is invisible to the user. 

The actual configuration used for the attack in our lab is as follows. Computer A and Computer B were the victims and Computer C acted as the attacker.
\medskip ~\\
\begin{minipage}{0.5\textwidth}
\small
\begin{tabular}[h]{rl}
Computer A: & Desktop computer \\
		  & (IP address: 172.20.122.84) \\ 
		  & Ethernet adapter \\
Computer B: & Desktop computer \\
		  & (IP address: 172.20.122.57) \\
		  & 3 Com 11Mbps Office Wireless Adapter\\                     
Computer C: & Laptop \\ 
		  & (IP address 172.20.122.114)\\ 
		  & VT6105 Rhine III Fast Ethernet Adapter \\
		  & Software used -- Cain and Abel, Ethereal
\end{tabular}
\end{minipage}

\begin{figure}
	\centering
		\includegraphics[width=7cm]{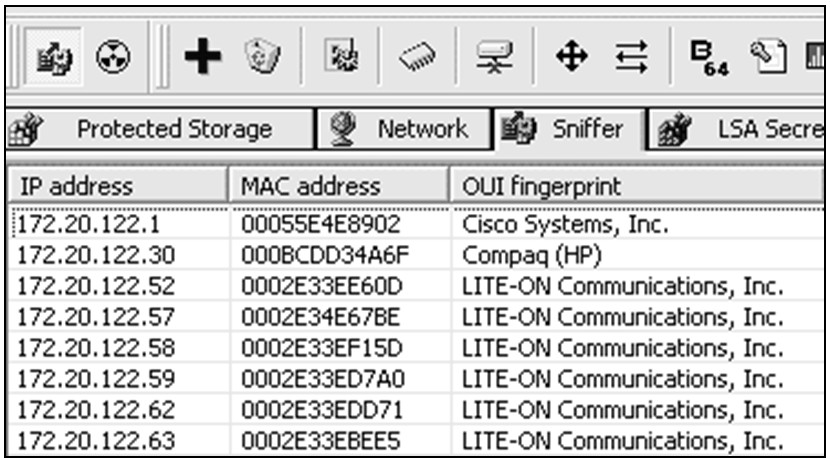}
		\caption{List of hosts in the network is shown using Cain and Abel software.}
	\label{fig:figure4}
\end{figure}

\medskip ~\\
Initially at Computer B, the ping command was given continuously to Computer A by issuing ``ping --t 172.20.122.84'' at the command prompt. At Computer C, Ethereal was initiated and the appropriate Ethernet adapter was selected. Computer C could only capture packets that were to and from it and broadcast packets. Thus Cain and Abel (ARP poisoning) software was needed to redirect the traffic to Computer C, so that it could sniff the ICMP packets. Without Cain and Abel, no ICMP packets would be shown in the Ethereal software (on C) as Computer B was pinging Computer A.

\begin{figure}
	\centering
		\includegraphics[width=7cm]{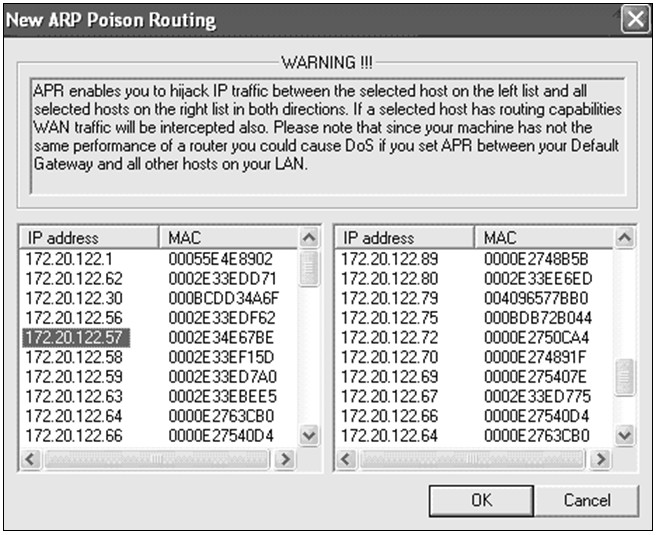}
	\caption{Selection of target computers to perform ARP poisoning in Cain and Abel software}
	\label{fig:figure5}
\end{figure}

Cain and Abel software was started and the appropriate adapter based on IP address is selected. Next the sniffer was started and all the hosts were selected for MAC address scanning. Then the lists of hosts available in the network would be displayed on Cain and Abel software.

The ARP tab was clicked and `` \textbf{+} '' tab is selected to add the victim computers to be poisoned. Computer B's IP was selected at the left column of the ``New ARP Poison Routing'' pop up window and Computer A's IP at the right column was also selected as in Figure 5. The start ARP tab was clicked to start the poisoning. The status ``Idle'' would change to ``Poisoning'' and packets transfer between Computer A and Computer B could be noticed on Computer C (attacker). Using Ethereal on Computer C, the ICMP packets transfer between Computer A and Computer B are shown in Figure 6. The communication between A and B can be monitored through Cain and Abel software as shown in Figure 7.

\begin{figure}
	\centering
		\includegraphics[width=8cm]{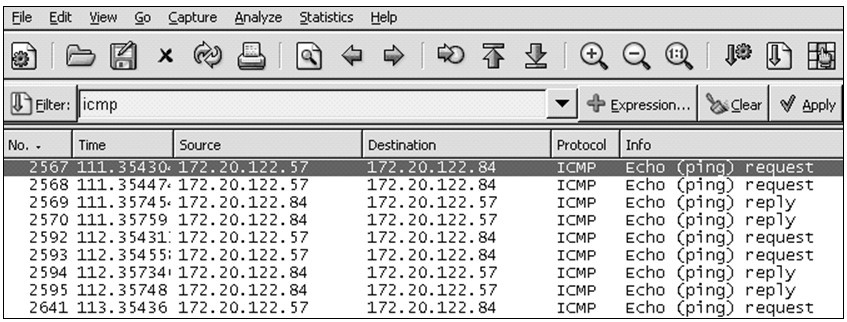}
	\caption{ICMP packets transferred between the target computers shown in Ethereal.}
	\label{fig:figure6}
\end{figure}

\begin{figure}
	\centering
\includegraphics[width=8cm]{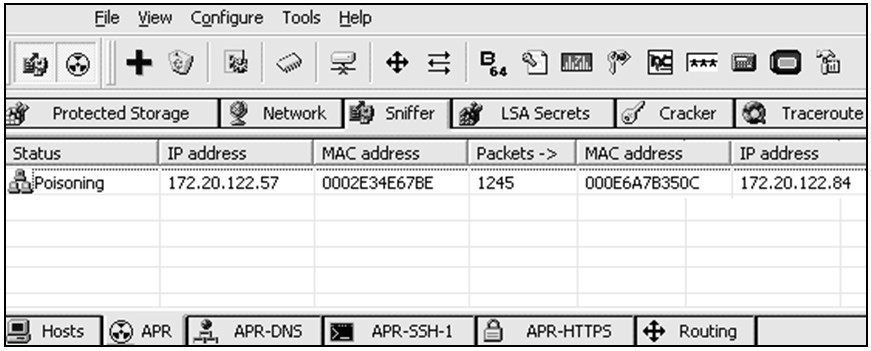}
\caption{ICMP packets transferred between A and B is monitored by Cain and Abel software.
}
\label{fig:figure7}
\end{figure}

\section{Related Work on Secure ARP}

A research publication [9] on Secure ARP (S-ARP) has been done by D. Bruschi et al. which deals with ARP broadcast communication security. Here each host has a public/private key pair certified by a local trusted party on the LAN, which acts as a Certification Authority. Messages are digitally signed by the sender, thus preventing the injection of spurious and/or spoofed information. It has been implemented also in Linux [9]. Tripunitara et. al. had outlined a middleware approach to the prevention of ARP cache poisoning as given in [10].

\section{The  Secure Unicast ARP (S-UARP) Protocol}
The S-UARP proposal we make is unicast in nature and have different options for security implementation. Many organizations would have implemented a DHCP server for dynamic IP address assignment to individual machines in a LAN. Hence the DHCP server can be configured to have the MAC-to-IP address mapping or vice-versa for all the computers/hosts under its domain.  We propose to extend the DHCP protocol to handle Secure Unicast Address Resolution Protocol (S-UARP) packets. We denote such a server as DHCP+ server from now on. The DHCP relay agent also needs to be modified to forward the S-UARP request/response messages. When using dynamic IP addressing using DHCP, the DHCP+ server stores the mapping of IP to MAC address as it leases out the IP address to the requesting hosts. We are not dealing with static IP addressing option in this section. But some suitable modification to this protocol can make it suitable for static addressing as noted in the next section. The proposal itself has an inherent partial-security against eavesdropping compared to ARP broadcast in a wired network, since packets are unicast in nature and is not broadcasted. In a wireless network, a packet sniffer can capture these unicast packets too since the radio transmission has no defined boundaries of transmission. But we add security into our protocol proposal.

\subsection{S-UARP Protocol}

This is a centralized protocol unlike the decentralized approach in normal ARP. Consider the following notations and their meaning as shown below.
\medskip ~\\
\small
\begin{tabular}{lrp{5cm}}
\centering
{\textbf Notation}& & {\textbf Meaning} \\
S-UARP\_req&: 	&S-UARP Request Packet \\
S-UARP\_res&:	&S-UARP Response Packet\\
DHCP+&: 		&DHCP+ Server\\
ICP&:		&Integrity Check Pass (security flag)\\
ICF&: 		&Integrity Check Fail (security flag)\\
A&:			&Host A\\
B&: 			&Host B\\
IP\_A&:		&IP address of A\\
MAC\_A&:		&MAC address of A\\
IP\_B&:		&IP address of B\\
MAC\_B&:		&MAC address of B\\
S$_{K}$&: 			&Session key \\
K$_{SA}$&:  		&Shared secret key between host A and the server \\
MIC&:			&Message Integrity Code\\
H&:			&Collision Free One-Way Hash Function\\
t&: 			&Time (independent variable) with one or more independent values.\\
t1&: 			&Time period (duration) when receiver waits for S-UARP\_req\\
t2&: 			&Time period when sender looks for a packet to be sent to the same host where ACK has to be sent.\\
t3&: 			&Time period within ACK packet has to be sent. (t3 > t2)\\
t4&: 			&Time period after which S-UARP cache needs refreshing.\\
\end{tabular}
\medskip ~\\
\normalsize

The S-UARP protocol (for dynamic IP addressing) is described as follows in 3 steps: 
\begin{enumerate}
\item[1.] A $\rightarrow$ DHCP+: S-UARP\_req
\item[2.] DHCP+  $\rightarrow$ A: S-UARP\_res + MIC  
\item[3.] A $\rightarrow$ DHCP+: (ACK)K$_{SA}$
\end{enumerate}

A simple example and explanation to show how this can be implemented with DES algorithm is as follows:

\begin{enumerate}
\item When a host \textit{A} wants to communicate to host \textit{B}, it sends a S-UARP request packet (unicast packet) to the DHCP+ server (which answers the S-UARP packets), instead of sending a broadcast to all. We assume that the secret hashing key (K$_{SA}$) is distributed between the client and the server, using private-public key mechanism or any other secure mechanism. 
\item The DHCP+ server encrypts the response message using DES with cipher block chaining (CBC). It cuts the message (S-UARP\_res) into predetermined-sized of \textit{i }blocks (where \textit{i }= \textit{1, 2, \ldots ., n}). Use the CBC residue (that is the last block output by CBC process) as a message integrity code (MIC). This MIC would act as a checksum [11]. The plaintext message plus the MIC would be transmitted to the host (receiver) or A. i.e. DHCP+ Server  Host A: Transmit S-UARP response (plain text) + MIC. The transmitted response message will be as in Figure 8.

\begin{figure}[!h]
	\centering
		\includegraphics[width=5cm]{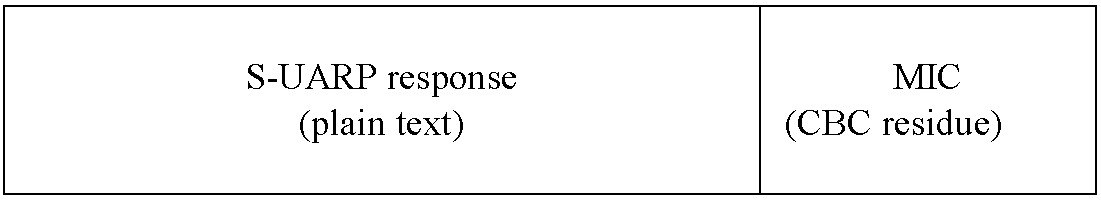}
	\caption{The S-UARP response message and MIC transmitted from DHCP+ Server. }
	\label{fig:figure8}
\end{figure}

If the response message doesn't arrive within a time period \textit{t}\textit{1}, host A will retransmit another S-UARP request packet to server.  This can continue until it gets a request packet.

\item Once the UARP response is received, host A checks for validity by using its secret key. The receiver (Host A) encrypts the plaintext S-UARP\_res using DES that it received with the shared secret key and do the hashing process to produce similar MIC (say, MIC*). Finally it checks the CBC residue or MIC. If MIC = MIC*, the message is a non-tampered in transit. We then call it Integrity Check Pass (ICP) state. Otherwise it is Integrity Check Fail (ICF) state and is discarded. The S-UARP response contains time t$_{s}$ when it was generated by the server. Host A also checks the freshness of the response by checking t$_{r}$ -- t$_{s}$ = $\Delta$t (similar to t3), where t$_{r}$ is the time when A receives the response from the server and $\Delta$t is the accepted time interval for transmission delay. Finally, the host A sends an encrypted acknowledgment (ACK)K$_{SA}$ to the server. ACK contains the timestamp \textit{t}\textit{$_{a}$} generated by the host A to ensure that the message is fresh and is not a replay. 
\end{enumerate}

The entries in S-UARP cache remains valid for a time period, \textit{t}\textit{4} (say, in minutes) as in ARP protocol. Once that time period expires, a new S-UARP request need to be sent by a host to DHCP+ server to get the IP-to-MAC address mapping. This can deal with a situation of change in ethernet card for a machine.

\subsection{Detailed Explanation}
The protocol can be shown in detail as follows, with possible optimization (as explained under section 5.3). When DHCP+ Server assigns a dynamic IP address to a host, the IP and MAC address of the DHCP+ server should be made known to the host.
\medskip ~\\
\small
\textit{Procedure} S-UARP\_Communication (A$\rightarrow$B)

BEGIN:

Initialize the\textit{ flag} [pkt\_send (from$\rightarrow$to)] = failure;

\textit{while}\textit{ }(pkt\_send (A$\rightarrow$DHCP+)  == failure) 

\{

\indent\indent Initialize t;

\indent\indent S-UARP\_req (IP\_A, MAC\_A, IP\_B);

\indent\indent A$\rightarrow$DHCP+: Sends S-UARP\_req;   //no broadcast

\indent\indent\textit{if (}$t < t1$)

\indent\indent\indent pkt\_send (A$\rightarrow$DHCP+) = success;

\indent\indent\textit{else}

\indent\indent\indent pkt\_send (A$\rightarrow$DHCP+) = failure;

\} //while loop

\textit{while } (pkt\_send (A$\rightarrow$DHCP+) == success $||$ $t > t3$)

\{

\indent\indent Initialize t;

\indent\indent S-UARP\_res (IP\_A, MAC\_A, IP\_B, MAC\_B, t$_{s}$)

\indent\indent DHCP+$\rightarrow$A: Sends UARP\_res + MIC;

\indent\indent\textit{if}  (pkt\_send (DHCP+$\rightarrow$A)  == success \&\& $t < t2$ \&\& 

\indent\indent\ ICP)

\indent\indent \{

\indent\indent\indent Host A$\rightarrow$DHCP+: Piggyback (ACK)K$_{SA}$;

\indent\indent\indent\textit{if} (pkt\_send (A$\rightarrow$DHCP+)  == success)

\indent\indent\indent\indent S-UARP Cache updated;

\indent\indent\indent\textit{else}

\indent\indent\indent\indent Go to start of enclosed \textit{while} loop; \textit{flag} = success;

\indent\indent\indent A$\rightarrow$B: A communicates to B directly;

\indent\indent \}

\indent\indent \textit{else if} (pkt\_send (DHCP+$\rightarrow$A) == success \&\& 

\indent\indent\ $t2 < t < t3$ \&\& ICP)

\indent\indent \{

\indent\indent\indent Host A  DHCP+: Sends (ACK)K$_{SA}$ packet;

\indent\indent\indent\textit{if} (pkt\_send (A$\rightarrow$DHCP+)  == success)

\indent\indent\indent\indent S-UARP Cache updated;

\indent\indent\indent\textit{else}

\indent\indent\indent\indent Go to start of enclosed \textit{while} loop; \textit{flag} = success; 

\indent\indent\indent A$\rightarrow$B: A communicates to B directly;

\indent\indent\}

\indent\indent\textit{else if} (pkt\_send (DHCP+$\rightarrow$A) == failure $||$ $t > t3$)

\indent\indent\{

   \indent\indent\indent Go to start of enclosed \textit{while} loop;

\indent\indent\}

\} //while loop

\textit{if}\textit{ } ($t > t4$ $||$ ICF )

\indent\indent S-UARP\_Communication (A$\rightarrow$B);

   END:  //end of procedure

\normalsize
\subsection{Possible Optimization}
An optimization possible is that the ACK can be piggybacked on another packet to the DHCP+ server, if packet transmission from host A to server happens within time $t2$. This can eliminate the separate ACK packet sent and save ACK congestion in the network. If there is no scope for piggybacking, and the acknowledgement is not received within a reasonable time period $t3$ (where $t3 > t2$), the server sends the S-UARP response packet again. If the S-UARP response packet is received by the host and the ACK packet is lost on transit, the duplicate response packets send by the server (after timeout $t3$) would be rejected. 

\subsection{Flow Chart for S-UARP}
The flow chart for the S-UARP protocol can be shown as in Figure 9.  It depicts the scenario when Host A wants to communicate to Host B (or a general Host X) and how the protocol works with respect to different time durations.

Note in Figure 9, $t1$ is the maximum time period for S-UARP response arrival (if it fails, host A would send another request), $t2$ is the maximum wait time for sending piggybacked ACK, t3 is the maximum acknowledgement wait time for sending ACK packet, where $t3 > t2$ (if $t > t3$, the server would send response again) and $t4$ is the maximum wait time, until S-UARP cache is refreshed.

\subsection{Alternate S-UARP Protocols (with more security)}

One of the limitations of the above protocol is that the request and the response are both in clear, though this is not a serious threat considering the content of the packets. Moreover, the message integrity is only on the server's response side. 

\textbf{\textit{Alternate Version 1:}} A better approach needs to ensure the integrity of both S-UARP request and response as follows:

\begin{enumerate}
\item[1.] A $\rightarrow$ DHCP+ : S-UARP\_req + MIC1
\item[2.] DHCP+ $\rightarrow$  A : S-UARP\_res + MIC2
\item[3.] A $\rightarrow$  DHCP+ : (ACK, NRN) K$_{SA}$
\end{enumerate}

In this protocol, we assume that a random number RN is known to both host and the server and is kept secret (generated by A or DHCP+). In step 1, A sends the request in clear and the MIC (i.e. MIC1). The MIC1 is generated using a collision-free one-way hash function like SHA1 that takes the secret key K$_{SA}$$_{,}$, the S-UARP\_req and the random number RN as inputs. That means, MIC1 = H(K$_{SA}$, RN, S-UARP\_req). In step 2, the server uses the S-UARP\_req (in plain text), the known random number RN and secret key, K$_{SA}$ to create a similar MIC (say, MIC1*). If MIC1 = MIC1*, then the request is accepted else it will be rejected. After verifying the integrity of the message, the server sends the response and MIC2 to the host. The MIC2 is generated in the same way (i.e. MIC2 = H(K$_{SA}$, RN, S-UARP\_res). Finally in step 3, Host A will check the integrity of the response as in the above case (to see MIC2 = MIC2*). Host A then sends an acknowledgement and a new random number (NRN) encrypted by the secret key (K$_{SA}$). NRN can be used in the next request/response exchange. As in the first protocol, the acknowledgment contains the timestamp to check when the server sent the response to the host, thus protecting against replay attacks.

\begin{figure}[!ht]
	\centering
		\includegraphics[width=7cm]{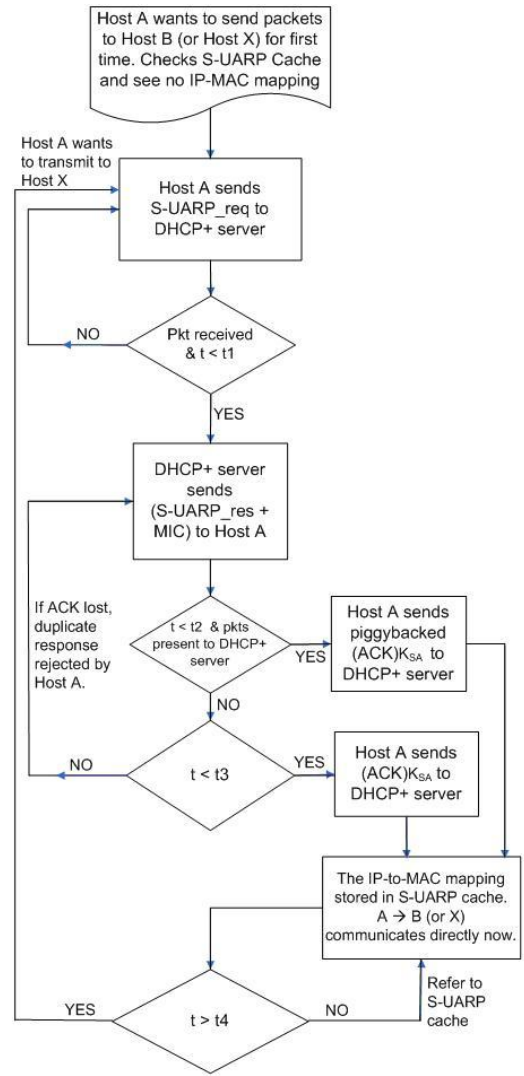}
	\caption{The flowchart showing the procedure of S-UARP operation. }
	\label{fig:figure9}
\end{figure}

\textbf{\textit{Alternate Version 2:}} Another more secure alternative is to use a session key \textit{S}\textit{$_{K}$}\textit{ }and an Exclusive-OR (XOR) operation as follows:

\begin{enumerate}
\item[1.] A $\rightarrow$ DHCP+ : S-UARP\_req + MIC1
\item[2.] DHCP+ $\rightarrow$ A : S- UARP\_res + SK  $\oplus$ MIC2 + MIC3
\item[3.] A  $\rightarrow$ DHCP+	: MIC4
\end{enumerate}

Here, MIC1 = H(K$_{SA}$, RN, S-UARP\_req), MIC2 = H(K$_{SA}$, S-UARP\_req\textit{,} S-UARP\_res), MIC3 = H(S$_{K}$, NRN), and MIC4 = H(S$_{K}$, ACK, NRN). In this protocol, the RN is generated by the server and is also known to host as a secret. In step 1,\textit{ A} sends the request and the MIC1 (using the key K$_{SA}$, RN and S-UARP\_req). In step 2, the server checks the integrity of the message (as shown in the previous protocols), and sends S-$_{ }$UARP\_res, S$_{K }$$_{}$$_{}$${\oplus}$$_{}$$_{ }$MIC2 and MIC3 to \textit{A}\textit{. }MIC2 and MIC3 are generated using the secret key and the session key respectively. MIC2 is XORed with session key, S$_{K}$$_{.}$  In step 3, host \textit{A} checks the integrity of the message received and then compute the acknowledgment as shown in MIC4. This acknowledgement calculation involves the timestamp as in previous cases. The NRN (generated by A or DHCP+) is used by the server in MIC3 is also contained in MIC4 and is kept secret by both parties for the next request/response exchange. It is clear here that even when an attacker knows \textit{K}\textit{$_{SA}$}, he will not be able to send the acknowledgment or MIC4 as he does not know the \textit{S}\textit{$_{K}$}\textit{$_{,}$}\textit{$_{ }$}used. As in the previous protocol, the attacker cannot also reply an old message (replay attack) since the ACK contains the timestamp when the server generated the message in step 2. It should be noted here that in all the three protocols, both requests and responses were sent in clear to avoid extra encryption overhead. The main objective is to ensure that the message was not modified in transit and to block the possibility of an ARP poisoning by an attacker.

\section{Issue of Static IP Addressing on Hosts}
In the previous section, we had ignored the issue of having static IP addressing on hosts, where we discussed the secure protocol with respect to dynamic IP addressing. There is still a need for central server (DHCP-) when we use static IP addressing, as follows:

The S-UARP protocol (for static IP addressing) is described as follows in the 3 steps: 
\begin{enumerate}
\item[1.] A $\rightarrow$ DHCP- : S-UARP\_req
\item[2.] DHCP-  $\rightarrow$ A : S-UARP\_res + MIC 
\item[3.] A   $\rightarrow$ DHCP- : (ACK)K$_{SA}$
\end{enumerate}

When the clients power up they advertise their encrypted IP address and MAC address to a central server (DHCP-), using the symmetric key. The DHCP- server keeps a record of the IP address and MAC address of all hosts in that network, much like a DHCP server, but doesn't issue IP addresses. It advertises its identity on a frequent basis and this takes precedence over normal DHCP addressing (if any) and clients would know whom to contact during ARP request. The ARP requests would then go unicast to DHCP- server from the clients as shown before.

\section{Co-existence of DHCP and DHCP+ Servers}

The DHCP+ server is an `improved' implementation of the normal DHCP server where it allows all DHCP queries to be directed to itself with security options, than doing a DHCP request broadcast as it is done in normal networks. If there is a situation where DHCP+ server implementation and DHCP server exist in the same network, the DHCP server needs to be patched to allow priority to DHCP+ server, so that the IP address assignment would only be done by the DHCP+ server. The software patching can help resolve the conflict of operation between the two within a same network. So in a co-existence scenario, the normal DHCP server would resign to a `passive' mode and DHCP+ server would be in an `active' mode.

\section{Mac Spoofing Attack and Implementation}
Mac spoofing is done where an attacker alters the manufacturer-assigned MAC address to any other value by using softwares like Mac Makeup[12] and [13] shows details of such attack. 

\begin{figure}[!ht]
	\centering
	\includegraphics[width=7cm]{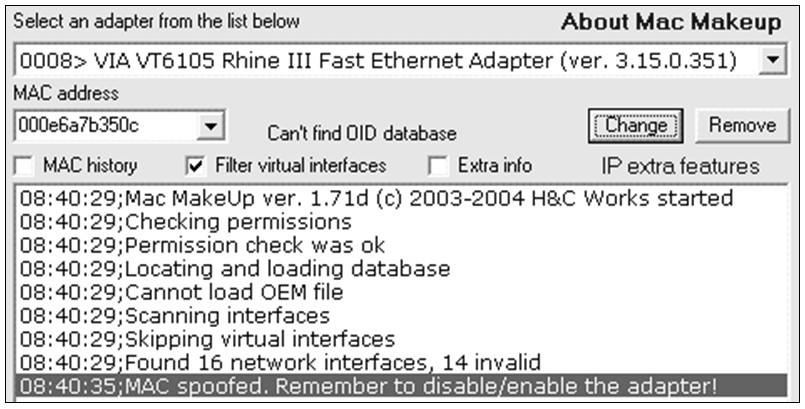}
	\caption{Mac Makeup software used to perform MAC spoofing for Ethernet adapter}
	\label{fig:figure10}
\end{figure}

In a brief experiment on a wireless LAN, the MAC address of the Intel® PRO/Wireless LAN 2100 3B Mini PCI Adapter that we were using was changed to the MAC address of a 3Com ll Mbps USB wireless adapter that was connected to the wireless network. Now it was found that the MAC address (0006a7b350c) and the IP address (172.20.122.88) assigned to the Intel Wireless Adapter is identical to the 3Com adapter.  

\begin{figure}[!h]
	\centering
		\includegraphics[width=5cm]{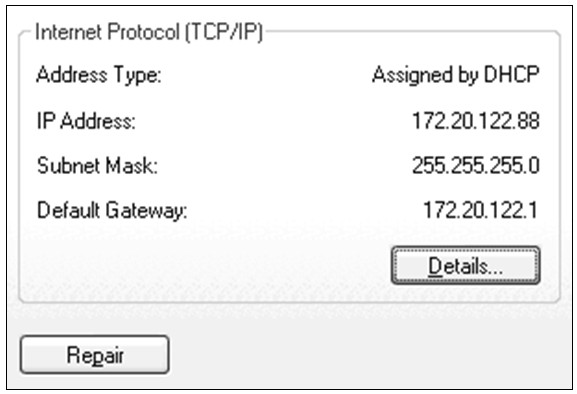}
	\caption{The IP address assigned to the attacker's Ethernet card}
	\label{fig:figure11}
\end{figure}

After identifying a MAC address to be spoofed, well-published DoS attack against the target was launched to cause the target's terminal to crash. In a real life attack, the attacker shall then immediately change the MAC address, IP address and default gateway to the value the target was using. With the target's computer rebooting, the attacker can access network resources bypassing the WLAN security appliance.

After MAC spoofing using Mac Makeup software as shown in Figure 10, the attacker's Ethernet card showed details as in Figure 11 and the overall network connection details as in Figure 12.

\begin{figure} [!ht]
	\centering
		\includegraphics[width=6cm]{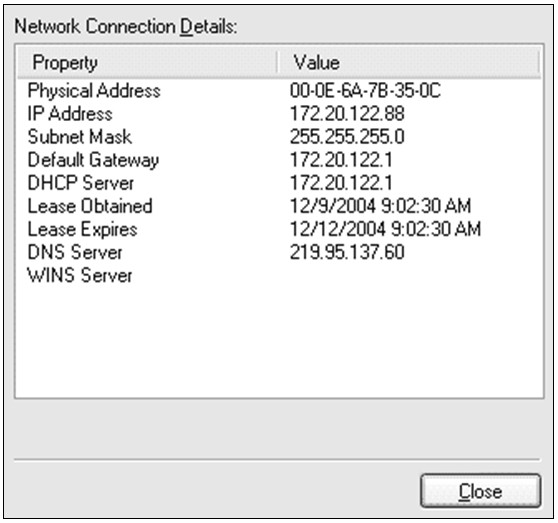}
	\caption{The spoofed IP and MAC address shown at attacker's computer}
	\label{fig:figure12}
\end{figure}

\begin{figure} [!ht]
	\centering
		\includegraphics[width=8cm]{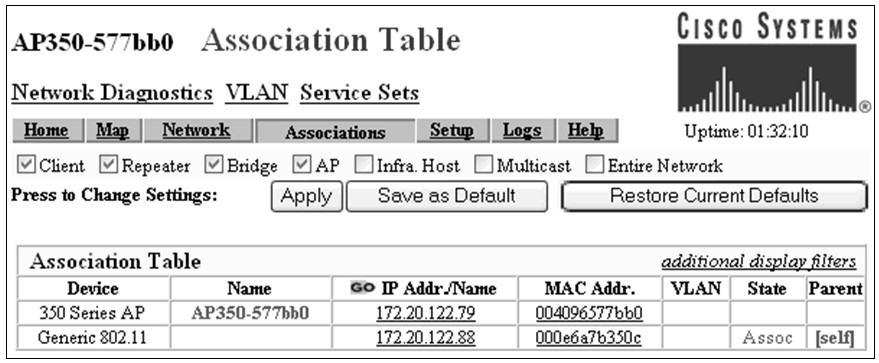}
	\caption{The spoofed IP and MAC address shown on Access Point's Association Table}
	\label{fig:figure13}
\end{figure}

In our case, the presence of firewall in Cisco AP was bypassed when the Ethernet card was used to spoof the MAC address of the wireless adapter and the Internet was browsed on a computer that used spoofed address. The access point's association table showed that the attacker's computer using spoofed MAC address was connected to wireless network as shown in figure 13.

\section{Secure DHCP Protocol}
The dynamic address assignment that is done by DHCP server also needs to be secured, against hacking.  MAC spoofing can easily be done through software to alter MAC addresses. We propose a secure DHCP (S-DHCP) protocol to make it less prone to MAC spoofing attacks.  Related work was done by Komori and Saito in [14] and another in [15]. Even if the MAC address is spoofed, the secure DHCP server will not assign the IP address, without proper credentials, as shown below. They don't take any additional steps than usual DHCP, except for the MIC overhead. The S-DHCP protocol is described as follows in 4 steps: 

\begin{enumerate}
\item[1.] A $\rightarrow$ ALL: Broadcast S-DHCP\_DISCOVER
\item[2.] DHCP+ $\rightarrow$ A: S-DHCP\_OFFER + (MIC1)K$_{SA}$ (unicast)
\item[3.] A $\rightarrow$ DHCP+: S-DHCP\_REQUEST + MIC1 + MIC2, to all DHCP servers that responded.
\item[4.] DHCP+ $\rightarrow$ A: S-DHCP\_ACK (unicast)
\end{enumerate}

Explanation of the secure version is generally similar to that for secure ARP. Host A broadcast a normal S-DHCP\_DISCOVER message packet. The DHCP+ server responds with a unicast secure S-DHCP\_OFFER message (that contains the IP address) appended with an encrypted Message Integrity Code (MIC1) using K$_{SA}$. K$_{SA}$ is the shared secret key between host A and DHCP+ server. MIC1 can be the CBC residue that is derived using DES CBC encryption method or the like, as outlined before under secure ARP. Host A would verify this message, by doing the same operation on the message and checks the result with MIC1 to see if it is same. Host A then responds by sending a secure S-DHCP\_REQUEST message appended with MIC1 and MIC2. MIC2 can be the CBC residue from encrypting S-DHCP\_REQUEST. DHCP+ server verifies this and sends a unicast acknowledgment S-DHCP\_ACK to host A. ACK contains the timestamp \textit{t}\textit{$_{s}$} generated by the server to ensure that the message is fresh and is not a replay. S-DHCP\_ACK can be optionally encrypted with K$_{SA}$. Only when DHCP+ server issues the ACK (step 4) that the IP address to client would be confirmed.

\begin{table*} 
\caption{Details of ARP packets in captured files}	\label{tab:01}
 \normalsize
\begin{center}
\begin{tabular}{|c|c|c|c|c|c|}\hline
Session &	No. of      & No. of &	No. of & Avg. ARP   & \% of ARP \\ 
No.     &	 hosts in   & total  & ARP     & packet size & pkts     \\
        &	 n/w        & pkts   & pkts    & (bytes) &               \\ [.5ex]
\hline
1&	48&	28366&	1326& 51.67&	4.67 \\
2&	45&	15539&	656&	59.15&	4.22 \\
3&	45&	10331&	557&	59.02&	5.39\\
4&	46&	15298&	650&	59.15&	4.25\\
5&	48&	12511&	668&	59.24&	5.34\\
6&	45&	17614&	677&	59.19&	3.84\\
7&	50&	11103&	646&	59.16&	5.82\\
8&	48&	16909&	675&	59.22&	3.99\\
9&	45&	11666&	583&	59.09&	5.00\\
10&	42&	11479&	562&	58.93&	4.90 \\ [.5ex]
\hline
\end{tabular}
\end{center}
\end{table*} 

\textbf{\textit{Alternate Version 1:}}  Another secure version of the protocol is shown as follows:

\begin{enumerate}
\item[1.] A $\rightarrow$ALL: Broadcast S-DHCP\_DISCOVER + MIC1
\item[2.] DHCP+ $\rightarrow$A: S-DHCP\_OFFER + MIC2 (unicast)
\item[3.] A $\rightarrow$DHCP+: S-DHCP\_REQUEST + MIC3, to all DHCP servers that responded.
\item[4.] DHCP+ $\rightarrow$A: (S-DHCP\_ACK, NRN) K$_{SA}$ (unicast)
\end{enumerate}

Where, MIC1= H(K$_{SA}$, RN, S-DHCP\_ DISCOVER), MIC2 = H(K$_{SA}$, RN, S-DHCP\_OFFER) and MIC3 = H(K$_{SA}$, RN, S-DHCP\_REQUEST). Explanation of the above protocol is similar to that has been done for secure ARP. In this protocol, we assume that a random number RN is known to both host and the server and is kept secret (generated by A or DHCP+). In step 1, A broadcasts the request in clear and the MIC (i.e. MIC1). The MIC1 is generated using a collision-free one-way hash function like SHA1 that takes the secret key K$_{SA,}$, the S-DHCP\_ DISCOVER and the random number RN as inputs, as listed above. In step 2, the server uses the S-DHCP\_DISCOVER (in plain text), the known random number RN and secret key, K$_{SA}$ to create a similar MIC (say, MIC1*). If MIC1 = MIC1*, then the request is accepted or else it will be rejected. After verifying the integrity of the message, the server sends the response and MIC2 to the host. The MIC2 is generated in the same way and is shown above. Finally in step 3, Host A will check the integrity of the response as in the above case (to see MIC2 = MIC2*). Host A then sends S-DHCP\_REQUEST (in plain) along with MIC3, as in previous steps. Finally, the server would check the integrity of the message from A and sends an acknowledgement and a new random number (NRN) encrypted by the secret key (K$_{SA}$). NRN can be used in the next request/response exchange. 

\textbf{\textit{Alternate Version 2:}} Another more secure version of the protocol is given below.

\begin{enumerate}
\item[1.] A $\rightarrow$ ALL: Broadcast S-DHCP\_DISCOVER + MIC1
\item[2.] DHCP+ $\rightarrow$A: S-DHCP\_OFFER + S$_K$  $\oplus$ MIC2 + MIC3
\item[3.]	A $\rightarrow$ DHCP+: S-DHCP\_REQUEST+ S$_K$ $\oplus$   MIC4 + MIC5, to all DHCP servers that responded.
\item[4.]   DHCP+ $\rightarrow$A: MIC6
\end{enumerate}

Where, MIC1 = H(K$_{SA}$, RN, S-DHCP\_DISCOVER), MIC2 = H(K$_{SA}$, S-DHCP\_DISCOVER\textit{,} S-DHCP\_OFFER), MIC3 = H(S$_{K}$, NRN), MIC4 = H(K$_{SA}$, S-DHCP\_OFFER\textit{,} S-DHCP\_REQUEST), MIC5 = H(S$_{K}$, NRN) and MIC6 = H(S$_{K}$, S-DHCP\_ACK, NRN). In this protocol, the RN is generated by the server and is also known to host as a secret. In step 1,\textit{ A} broadcasts S-DHCP\_DISCOVER and the MIC1. In step 2, the server checks the integrity of the message (as shown in the previous protocols), and sends S-DHCP\_OFFER, S$_{K }$$_{}$$_{}$${\oplus}$$_{}$$_{ }$MIC2 and MIC3 to \textit{A. }MIC2 and MIC3 are generated using the secret key and the session key respectively. MIC2 is XORed with session key, S$_{K.}$  In step 3, host A checks the integrity of the message (as shown in the previous protocols), and sends S-DHCP\_REQUEST, S$_{K }$$_{}$$_{}$${\oplus}$$_{}$$_{ }$MIC4 and MIC5 to server\textit{. }MIC4 and MIC5 are generated using the secret key and the session key respectively. MIC4 is XORed with session key, S$_{K.}$$_{ }$In step 4, the server checks the integrity of the message received and then computes the acknowledgment as shown in MIC6. This acknowledgement calculation involves the timestamp as in previous cases. The NRN (generated by A or DHCP+) is used by the server in MIC3 (which is also contained in MIC5 and MIC6) is kept secret by both parties for the next request/response exchange. It is clear here that even when an attacker knows \textit{K}\textit{$_{SA}$}, he will not be able to send the acknowledgment or MIC4 as he does not know the \textit{S}\textit{$_{K, }$}used. As in the previous protocol, the attacker cannot also reply an old message (replay attack) since the ACK contains the timestamp when host A generated the message in step 3.

\section{Performance Analysis of ARP and DHCP Protocols}
We captured live packet traffic from a wired office network using Ethereal software [12] and filtered all the ARP packets out to do an analysis of ARP packets as in Figure 14. The traffic analysis below shows the percentage of ARP packets found in packet samples collected for around 30 minutes each during 10 sessions. This is shown in Table 1.  On the average the network contained around 40 to 50 hosts (clients and servers), including print servers.  There were five HP Jet-Direct Print Servers, three Canon Network Print Servers and one D-Link Print Server. The operating system platform was mostly Windows XP on clients, along with Linux server and Netware Server. The percentage of ARP packets was found to be around 4\% to 5\% of the total traffic.

\begin{table*}
\caption{ARP, S-UARP (no ACK) and SARP Comparison}\label{tab:02}
\normalsize
\begin{center}
\begin{tabular}{|c|c|c|c|c|c|c|}\hline
No. & No. of  & No. of    & No. of      &\% of ARP  &\% of     &\% of  \\ 
   &total    & ARP       & S-UARP pkts & pkts	     & S-UARP   &SARP   \\
   &pkts     & pkts      & [No ACK]    &   & pkts     &pkts   \\ [.5ex]
\hline
1&	28366&	1326&	228&	4.67&	0.80&	7.09\\
2&	15539&	656&	60&	4.22&	0.39&	5.38\\
3&	10331&	557&	62&	5.39&	0.60&	7.19\\
4&	15298&	650&	64&	4.25&	0.42&	5.50\\
5&	12511&	668&	58&	5.34&	0.46&	6.73\\
6&	17614&	677&	62&	3.84&	0.35&	4.90\\
7&	11103&	646&	62&	5.82&	0.56&	7.49\\
8&	16909&	675&	60&	3.99&	0.35&	5.06\\
9&	11666&	583&	60&	5.00&	0.51&	6.54\\
10&	11479&	562&	72&	4.90&	0.63&	6.78\\ [.5ex]
\hline
\multicolumn{7}{|c|}{Average Broadcast Packet Reduction in S-UARP (w.r.to ARP) 
= 09.77 times} \\ [.5ex]
\hline
\end{tabular}
\end{center}
\end{table*}

\begin{figure} [ht]
 \centering
	\includegraphics[width=7cm]{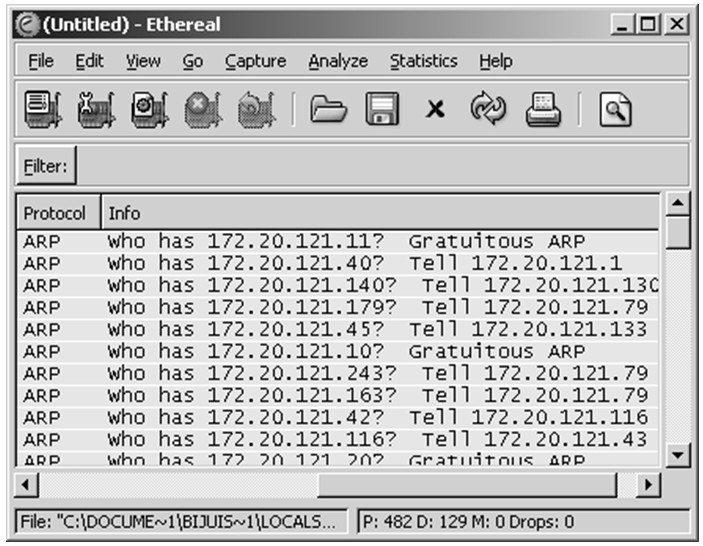}
	\caption{Sample of ARP Packet capture using Ethereal.}
	\label{fig:figure14}
\end{figure}

\begin{figure} [ht]
	\centering
		\includegraphics[width=7cm]{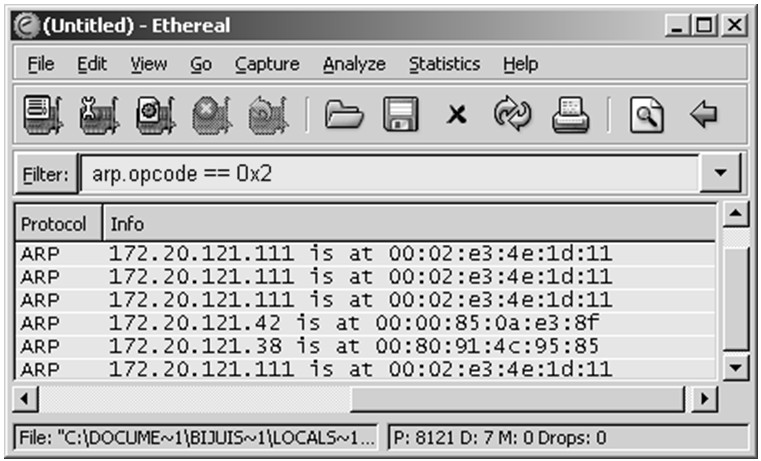}
	\label{fig:figure15}
	\caption{Sample of ARP Reply packets captured using Ethereal.}
\end{figure}

When we calculate S-UARP packet details, we assume that the channel is free of errors and there are no retransmissions required. Also we assume that the ACK is piggybacked every time. The Table 2 shows the Broadcast Packet Reduction because of S-UARP implementation. The S-UARP packet count is done by finding the number of ARP replies and multiplying that by 2. 

This is because S-UARP is unicast and hence there would only be 2 packets exchanged (request and reply) between host and server, excluding the ACK packets. So the calculation can be as follows: No. of S-UARP packets (no ACK) = 2 x No. of ARP reply packets. To get the ARP reply packets, we need to use the ethereal software. Ethereal filter can be enabled with the expression, arp.opcode == 0x2, which is the opcode for ARP reply packet, to get all the ARP reply packets as in Figure 15.

It's quite clear when the number of computers in the network increases the ARP broadcast can still be higher.

The results in Table 2 show that there is a reduction in unwanted broadcast packets by 9.77 times (excluding ACK packets, which is piggybacked). The value is an average of 10 samples. SARP is any other secure ARP scheme that uses PKI infrastructure that needs 4 steps to complete an ARP request cycle. The comparison graph can be as in Figure 16. 

The S-UARP channel link utilization with ACK packets is shown in Table 3. Here we assume the worst case of no piggybacking ACK. Thus ACK is sent as a separate packet. Again, we don't consider any retransmission cases here and assume that the channel is free of such errors. The no. of S-UARP packets (with ACK) = 3 x No. of ARP reply packets; As 3 packets are needed to be exchanged here for one cycle -- i.e. S-UARP request, S-UARP response and ACK.

\begin{figure} [!ht]
	\centering
		\includegraphics[width=7cm]{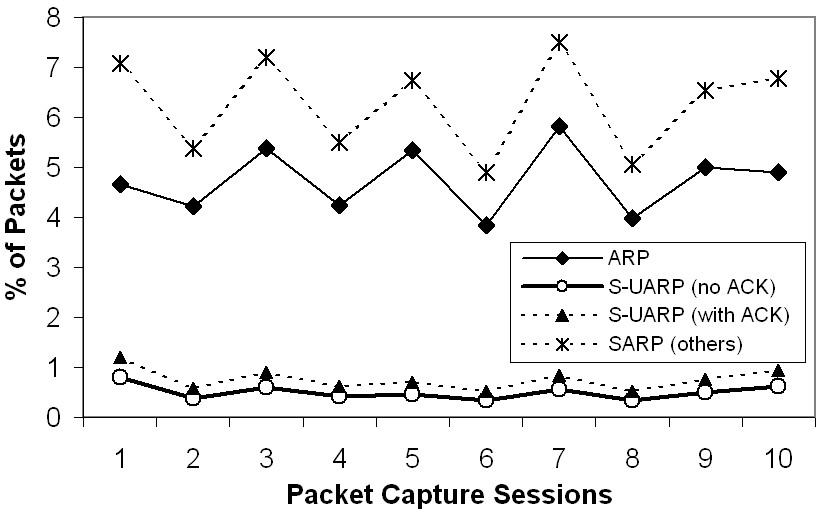}
	\caption{Host Channel Link Utilization (ARP, S-UARP without ACK, S-UARP with ACK and SARP).}
	\label{fig:figure16}
\end{figure}

There can be a 6.50 times reduction in congestion through S-UARP packets (with ACK) as seen in Table 3. The comparison line graph can be as shown in Figure 16. Assuming that, piggybacking can happen with ACK transmission (host to DHCP+ server) for about 50\% time, the Broadcast Packet Reduction can be around 8.13 times (average of previous two cases) than in a normal ARP scenario. This is quite a good result.

\begin{table*}
\caption{ARP, S-UARP (with ACK) and SARP Comparison} \label{tab:03}
\normalsize
\begin{center}
	\begin{tabular}{|c|c|c|c|c|c|c|}\hline
	
No. & No. of & No. of & No. of      &\% of ARP &\% of   &\% of \\
    & total  & ARP    & S-UARP pkts & pkts	 & S-UARP & SARP \\
    & pkts   & pkts   & [No ACK]    &  	 & pkts   & pkts \\ [.5ex]
\hline
1&	28366&	1326& 342& 4.67& 	1.21& 7.09\\ 
2&	15539&	656&	90&	4.22&	0.58&	5.38\\
3&	10331&	557&	93&	5.39&	0.90&	7.19\\
4&	15298&	650&	96&	4.25&	0.63&	5.50\\
5&	12511&	668&	87&	5.34&	0.70&	6.73\\
6&	17614&	677&	93&	3.84&	0.53&	4.90\\
7&	11103&	646&	93&	5.82&	0.84&	7.49\\
8&	16909&	675&	90&	3.99&	0.53&	5.06\\
9&	11666&	583&	90&	5.00&	0.77&	6.54\\
10&	11479&	562&	108&	4.90&	0.94&	6.78\\ [.5ex]
\hline
\multicolumn{7}{|c|}{Average Broadcast Packet Reduction in S-UARP (w.r.to ARP) 
= 6.50 times} \\ [.5ex]
\hline

\end{tabular}
\end{center}
\end{table*}

\begin{figure} [!ht]
	\centering
	\includegraphics[width=7cm]{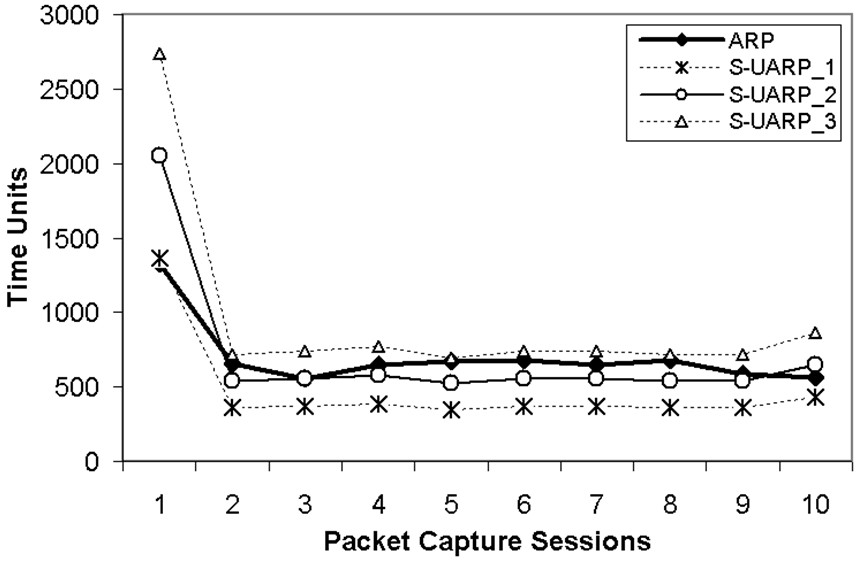}
	\caption{Overall time consumed per session for each of the ARP schemes (ARP vs. three S-UARP versions).}
	\label{fig:figure17}
\end{figure}

Considering that the encryption operation to carry a factor of 2, compared to a normal operation, the time consumption for the different ARP schemes would be as shown in Figure 17. Here any step that uses encryption (considering the encryption and decryption process) is given double the weight of a normal step without encryption. S-UARP\_1, S-UARP\_2 and S-UARP\_3 are the three proposed schemes with increasing security. It shows SUARP scheme with light encryption is better in time than normal ARP. Figure 18 shows the graph, ignoring encryption performed on ACK.

\begin{figure} 
	\centering
	\includegraphics[width=7cm]{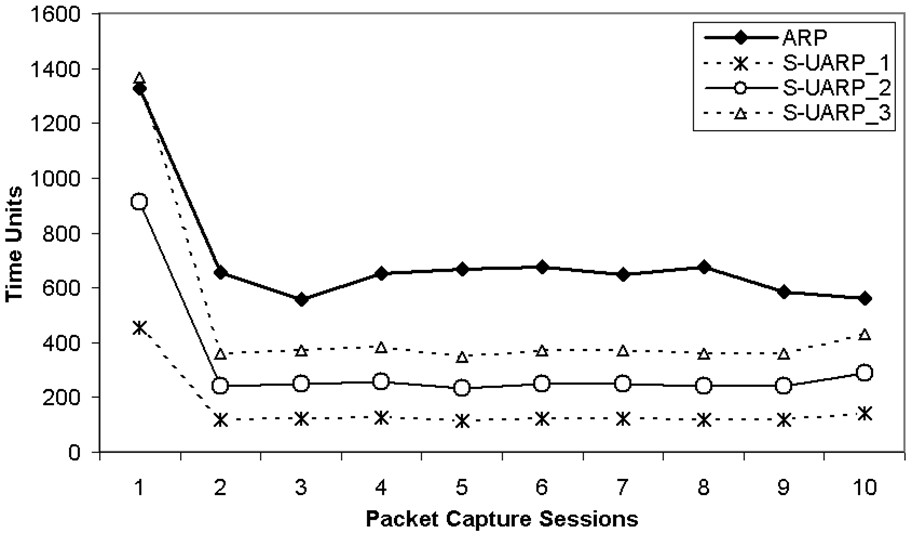}
	\caption{Overall time consumed per session for each of the ARP schemes (ARP vs. three S-UARP versions), ignoring the encryption performed for ACK sent.}
	\label{fig:figure18}
\end{figure}

Since secure DHCP(S-DHCP) uses the same number of steps as its original version, the only overhead encountered would be that of calculating MICs, appending them etc. Like before, a factor of 2 is assigned to all encryption steps. The Figure 19 shows the comparisons, for each of the 10 sessions for a random sample of packets. It is right to infer from the graph that the basic version with minimum security (S-DHCP version 1) has lesser computation overhead and lesser delay.

\begin{figure} [!ht]
	\centering
	\includegraphics[width=7cm]{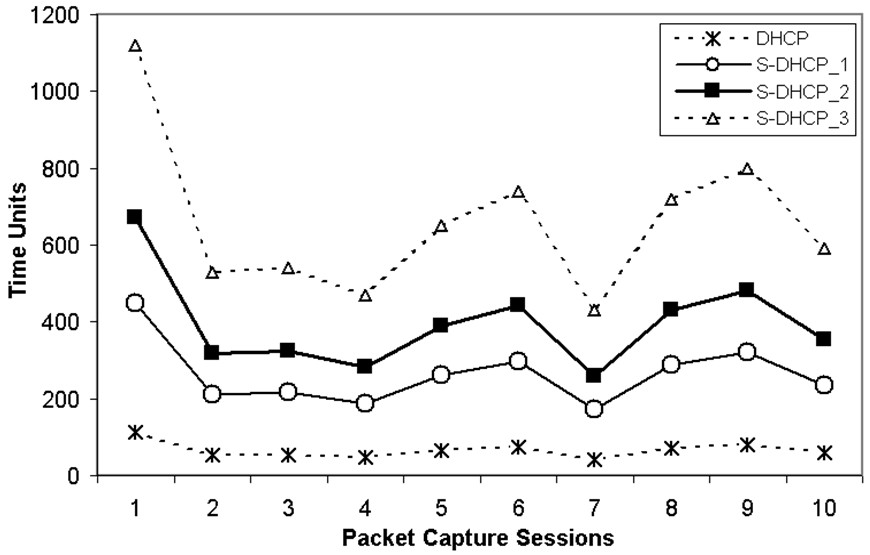}
	\caption{Overall time consumed per session for each of the DHCP schemes (DHCP vs. three S-DHCP versions).}  
	\label{fig:figure19}
\end{figure}

\section{Conclusion}
Though some initiatives had been there to mitigate ARP poisoning, the new S-UARP protocol (along with secure DHCP) is more efficient in terms of performance and security. It reduces broadcast congestion in network, since the S-UARP request is unicast and directed to only the secure DHCP server. It is quite difficult for an attacker to do ARP poisoning attack, especially on the more secure versions of S-UARP. It is thus protected against message integrity attacks (when ARP packet content can be modified by attacker) and masquerading attacks (when new ARP bogus packet injection can be done by attacker). Also since the DHCP protocol is made secure, the MAC spoofing attacks are also eliminated. The performance analysis of both the protocols are also discussed. This proposal is mostly relevant to IPv4 networks, since ARP is implemented only in IPv4 networks. IPv6 networks use a different mechanism (called Neighbor Discovery Protocol). Nevertheless it is quite relevant until a whole conversion to IPv6 from IPv4 fully happens.

\section*{Acknowledgments}
This paper is a major extension of the paper titled -- ``Secure Unicast Address Resolution Protocol (S-UARP) by Extending DHCP'' that was presented during ICON 2005, Malaysia.

\newenvironment{biography}[1]{%
\footnotesize\unitlength 1mm\bigskip\bigskip\bigskip\parskip=0pt\par%
\rule{0pt}{39mm}\vspace{-39mm}\par
\noindent\setbox0\hbox{\framebox(25,32){}}
\ht0=37mm\count10=\ht0\divide\count10 by\baselineskip
\global\hangindent29mm\global\hangafter-\count10%
\hskip-28.5mm\setbox0\hbox to 28.5mm {\raise-30.5mm\box0\hss}%
\dp0=0mm\ht0=0mm\box0\noindent\bf#1\rm}{
\par\rm\normalsize}

\begin{biography}
{Biju Issac} is a lecturer in Information Technology at Swinburne University of Technology (Sarawak Campus), Malaysia. He is holding a BEng (Electronics and Communication Engineering) degree along with an MCA (Master of Computer Application) with Honours from Calicut University, India. He is an IEEE, IEEE Communication Society and IEEE Education Society member. His research interests are mainly in computer networks and education. Specifically, his research interest is in mobility management, wireless and network security, education and e-learning. He is heading the network security research in Information and Security Research (iSECURES) Lab in Swinburne. He has a number of refereed publications that includes many conference papers, journal papers and book chapters.

\end{biography}


\begin{thebibliography}{10}
\bibitem{1}
A. S. Tanenbaum, \textit{Computer Networks}, 4th edition, Prentice Hall PTR, pp.450-454, 2003.
\bibitem{2}

D. C. Plummer, ``RFC 826 -ARP Protocol'', 1982. (http://www.faqs.org/rfcs/rfc826.html)
\bibitem{3}

C. M. Kozierok, The TCP/IP Guide Website, 2005. (http://www.tcpipguide.com/free/
t\_ARPMessageFormat.htm)
\bibitem{4}

R. Droms, ``RFC 2131 --Dynamic Host Configuration Protocol'', 1997. (http://www.faqs.org/rfcs/rfc2131.html)
\bibitem{5}

S. Alexander, ``RFC 2132 --DHCP Options and BOOTP Vendor Extensions'', 1997. (http://rfc.net/rfc2132.html)
\bibitem{6}

C. Nachreiner, "Anatomy of an ARP Poisoning Attack", Washington, USA, 2003. (http://www.watchguard.com/infocenter/editorial/
135324.asp)
\bibitem{7}

The Ethereal Software, version 0.99.0 (http://www.ethereal.com)
\bibitem{8}

The Cain \& Abel Software, version 2.5 (htp://www.oxidt.it and http://www.nwcet.org/
downloads/cainAbel.pdf)
\bibitem{9}

D. Bruschi, A. Ornaghi and E. Rosti, ``S-ARP: a Secure Address Resolution Protocol'', \textit{19th Annual Computer Security Applications Conference}, pp.66-74, Nevada, USA, 2003.
\bibitem{10}

M. V. Tripunitara and P. Dutta. A middleware approach to asynchronous and backward compatible detection and prevention of ARP cache poisoning, in the \textit{Proceedings of the 15th Annual Computer Security Application Conference (ACSAC)}, pp 303-309, 1999.
\bibitem{11}

Larry L. Peterson and Bruce S. Davie, \textit{Computer Networks -- A systems approach}, 3rd edition, pp.583-601, Morgan Kaufmann (Elsevier), 2003.
\bibitem{12}

The Mac Makeup Software, version 1.71d (http://www.gorlani.com/publicprj/MacMakeUp/
macmakeup.asp)
\bibitem{13}

J. Wright, Detecting Wireless LAN MAC Address Spoofing, Johnson and Wales University, GCIH, CCNA, 2003. (http://home.jwu.edu/jwright/papers/wlan-mac-spoof.pdf)
\bibitem{14}

T. Komori and T. Saito, The secure DHCP system with user authentication, in the \textit{Proceedings of the 27th Annual IEEE Conference on Local Computer Networks (LCN)}, pp. 123-131, Florida, USA, 2002.
\bibitem{15}

H. Altunbasak, S. Krasser, H. Owen, J. Sokol. and J. Grimminger, Addressing the weak link between layer 2 and layer 3 in the Internet architecture, in the \textit{Proceedings of 29th Annual IEEE International Conference on Local Computer Networks (LCN)}, pp.417-418, Florida, USA, 2004.
\end{thebibliography}
\end{document}